\documentclass[aps,preprint,showpacs,superscriptaddress,groupedaddress]{aastex}  
\usepackage{graphicx}
\usepackage{dcolumn}
\usepackage{bm}
\usepackage{amssymb}
\usepackage{graphicx}
\usepackage{dcolumn}
\usepackage{bm}
\usepackage{amsmath, amssymb, amsfonts, amsthm}

\newcommand\simlt{\lower.5ex\hbox{$\; \buildrel < \over \sim \;$}}
\newcommand\simgt{\lower.5ex\hbox{$\; \buildrel > \over \sim \;$}}

\begin{document}
\tighten
\title{Broadband extended emission in gravitational waves from core-collapse supernovae}
\author{Amir Levinson$^{a}$,  Maurice, H.P.M. Van Putten$^{b}$\footnote{corresponding author: mvp@sejong.ac.kr}, and Guy Pick$^a$  }
\affil{$^a$ Raymond and Beverly Sackler School of Physics and Astronomy\\ Tel Aviv University,
Tel Aviv 69978, Israel}
\affil{$^b$ Room 614, Astronomy and Space Science, Sejong University, 98 Gunja-Dong Gwangin-gu, Seoul 143-747, Korea}





\begin{abstract}
Black holes in core-collapse of massive stars are expected to surge in mass and angular momentum by hyper-accretion immediately following their formation. We here describe a general framework of extended emission in gravitational waves from non-axisymmetric accretion flows from fallback matter of the progenitor envelope. It shows (a) a maximum efficiency in conversion of accretion energy into gravitational waves at hyper-accretion rates exceeding a critical value set by the ratio of the quadrupole mass inhomogeneity and viscosity with (b) a peak characteristic strain amplitude at the frequency $f_b=\Omega_b/\pi$, where $\Omega_b$ is the Keplerian angular velocity at which viscous torques equal angular momentum loss in gravitational radiation, with $h_{char}\propto f^{1/6}$ at $f<f_b$ and $h_{char}\propto f^{-1/6}$ at $f>f_b$. Upcoming gravitational wave observations may probe this scaling by extracting broadband spectra using time-sliced matched filtering with chirp templates, recently developed for identifying turbulence in noisy time series.
\end{abstract} 

\maketitle

\section{Introduction}
The birth of stellar mass black holes in core-collapse supernovae of relatively massive stars is 
believed to be associated with hyper-accretion of fallback matter from the progenitor's envelope \citep{bet03}.   
The black hole hereby surges in mass and angular momentum \citep{bar70,van04} until matter from the envelope is exhausted or 
until the black hole reaches near-extremal spin perhaps close to the Thorne limit \citep{tho74}. The black hole may subsequently experience spin-down by strong interactions with high density matter at the ISCO via an inner torus magnetosphere as \citep{van99}.

This general outlook on accreting stellar mass black holes is receiving increasing attention as a leading candidate of the central engines of long GRBs and low luminosity long GRBs and, conceivably, the more extended class of broad-line hyper-energetic core collapse supernovae (CC-SNe). Although an origin in millisecond magnetars \citep{uso92,met11} cannot be ruled out, rapidly rotating black holes offer multimessenger emissions through frame dragging induced interactions by an ample energy reservoir in spin energy, that can readily account for the most extreme GRB-supernovae \citep{van11a}. By a diversity in spin, rotating black holes further offer a unification of short and long GRBs \citep{van01a}. 

   The focus of the present discussion is hereby complementary to and different from that in existing models for gravitational
   waves from CC-SNe producing (proto-)neutron stars. The complex dynamical process of core-collapse
   and core-bounce in the first one or two seconds at birth of a neutron star opens a remarkably broad window to
   gravitational radiation \citep{ott09} of potential interest to upcoming observations of events at distances
   up to a few Mpc \citep{rov09}. The connection to GRB-supernovae and extremely energetic supernovae, however, is 
   not obvious \citep{bur07,des08,moe15}.

Non-axisymmetric high-density accretion disks orbiting black holes are potentially prodigious emitters of gravitational radiation. 
A fraction of order unity of the gravitational binding energy of accreted matter released in gravitational waves would provide just the kind of energetic output sufficient for a detection within a distance on the order of $D\simeq 100$ Mpc by advanced gravitational wave detectors LIGO-Virgo and KAGRA using
Time Sliced Matched Filtering \citep[TSMF;][]{van11,van14,van15}. These emissions, if present, tend to produce gravitational wave chirps from mass-inhomogeneities in accretion flow, down to wave-instabilities about the ISCO in suspended accretion flows \citep{van01,van02}. The latter offers a window further to be powered by the spin energy of the black hole, giving a unique perspective on detecting black hole evolution. Ultimately, detection of gravitational wave emissions from a core-collapse supernova is expect to rigorously identify the nature of the inner engine of long GRBs \citep{cut02}, i.e., a black hole-disk system or magnetar (e.g. \cite{mel14}), that presently remains elusive based on electromagnetic observations alone.

Quite a few mechanisms can naturally give rise to non-axisymmetric accretion flows. First, the envelope of the progenitor is likely to be highly turbulent, leading to intermittent fallback of stellar material. Second, hydrodynamical instabilities give rise to unsteady accretion with large temporal variations of the accretion flow parameters  \citep{mac99}. In 3D, it is conceivable that these instabilities excite non-axisymmetric azimuthal modes. A leading mechanism for generating mass-inhomogeneities is a gravitational instability \citep{too64,gol65,ada89,had14}. Under idealized conditions, numerical simulations of thin disks \citep[e.g.,][]{gam01,ric05} indicate excitation of low-$m$ modes when self-gravity becomes important. 
It can even lead to disk fragmentation if the cooling rate in the instability zone is sufficiently high \citep{gam01,ric05,sad15}.
Although it remains to be demonstrated by direct numerical simulations, conceivably this mechanism applies also to collapsar disks at hyper-accretion rates relevant to long GRBs and some of the more energetic CC-SNe. Even without self-gravity, the accretion disk may also develop internal wave instabilities, such as spiral waves of Lin-Shu type  \citep[e.g.,][]{gri11} or Rossby waves driven by MHD stresses 
\citep{tag90,tag99,tag01,tag06,tag06b,lov14}. Third, Papaloizou-Pringle wave instabilities may be excited at the surface of an inner torus about the ISCO \citep{pap84,van02} provided accretion is sufficiently suppressed \citep{bla87}, e.g., by feedback from a rapidly rotating Kerr black hole by magnetic 
coupling \citep{van99,van03}. Fourth, accretion flows may be susceptible to transient shocks that may migrate over an extended range of radii, ingoing or
outgoing, with associated chirp-like behavior \citep{cha04,cha08} that conceivably break axisymmetry. Common to all these mechanisms is a gravitational wave output from mass-inhomogeneities in non-axisymmetic accretion flows.

Gravitational radiation from accretion disks has been discussed earlier in several papers, in connection with black hole spin \citep[][]{van02,van03}, extended accretion flows \citep{kob03,pir07}, and recently also by numerical simulations \citep{kor11,kiu11,mew15}.
For self-gravitating disks, the Toomre criterion has been employed to argue that a collapsar disk should become gravitationally unstable at radii $r/M>300 \alpha_{-1}^{2/3}\dot{M_1}^{-2/3}$ \citep{pir07}, where $\alpha=0.1\alpha_{-1}$ is the alpha viscosity parameter and  $\dot{M}=\dot{M}_{1}\ M_\odot s^{-1}$ is the accretion rate. Several cooling processes have been examined, including electron-positron pair annihilation to neutrinos, URCA process, and photo-disintegration  of $^4$He, with the latter found to be the most effective one in the instability zone. 
It has been argued \citep{pir07} that cooling by photo-disintegration may lead to fragmentation chirps in gravitationally bound clumps, migrating inwards with frequencies $0.1 -1 $ kHz depending on disk parameters and fragmentation mass. For LIGO, it suggests a sensitivity distance of about $100$ Mpc \cite{pir07}. What fraction of the total mass accreted, if at all, will end up as gravitationally bound clumps, and how many clumps are formed are yet unresolved issues.  

Prodigious gravitational radiation at twice the orbital frequency about the ISCO has been predicted in the suspended accretion scenario \citep{van03}. In this model, the major fraction of the  spin energy of a Kerr black hole is transferred to the surrounding torus via magnetic field lines anchored to it, and is emitted as  gravitational waves, MeV neutrinos and winds.  A minor fraction is released along open magnetic field lines in the polar region in the form of a ultra-relativistic jet. The torus is subject to a self-regulated instability, with a quadrupole 
mass inhomogeneity of the order of 10\% at saturation level.  It is worth noting that in this scenario, the energy source of the gravitational waves emitted by the torus is the rotational energy of the black hole, rather than the gravitational binding energy of the surrounding matter, as in the gravitational radiation from accretion flows.  The suspended accretion model requires a strong suppression of mass loading of magnetic field lines that connect the torus and the black hole \citep{glo13}, possibly supported by surface gravity in the inner face of the torus by its inherently super-Keplerian motion. It could be that this process may be inherently intermittent \citep{van99}. Observational support for the associated loss of angular momentum of the black hole, following
a near-extremal state formed in a prior epoch of hyper-accretion, is found in normalized BATSE light curves \citep{van12,van15}. 

In light of the various distinct physical processes by which non-axisymmetric accretion flows may originate in core-collapse events, 
we here consider a general framework for their gravitational wave emissions in hyper-accretion flows following black hole formation.
Our primary objective is to identify some canonical scaling behavior in broadband emissions from radially extended
non-axisymmetric features in accretion flows. Our focus is on relatively long-lasting features potentially lasting tens of seconds in
accretion flows derived from fall back matter from a progenitor stellar envelope, posterior to a relatively brief but 
complex transition in the formation and settling down of the inner accretion disks immediately following black hole formation, 
recently highlighted in detailed axisymmetric numerical simulations \citep{sek11}.

Specifically, we set out to identify the frequency at maximal gravitational wave luminosity following a transition between viscous and gravitational wave-driven accretion flows. These general results may be confirmed in future large scale numerical simulations of accretion flows, that take into
account gravitational wave emission and their back reaction on the accretion flow. 

\section{GWs from non-axisymmetric accretion flow}

We consider a mathematical discretization of radially extended non-axisymmetric features in accretion disks by 
annular rings, each rotating at the local Keplerian angular velocity
\begin{eqnarray}
\Omega(r)=\frac{1}{M}\left(\frac{M}{r}\right)^{3/2},
\label{EQN_OMK}
\end{eqnarray}
around a black hole of mass $M$. We henceforth use geometrical units, $G=c=1$, whereby mass accretion rate is measured in units of 
\begin{eqnarray}
\dot{m}_0=\frac{c^3}{G}=4\times10^{38}\,\mbox{ g s}^{-1},
\end{eqnarray}
and luminosity in units of $L_0=\dot{m}_0c^2$. 
The underlying non-axisymmetric features themselves may be continuous, as in spiral waves, or discrete, in case of lumps. 
In what follows, we shall develop a general frame work for the associated broadband gravitational wave emission based on
our ring discretization, that applies irrespective of the detailed properties these features. The condition for our ring
discretization to apply is rather mild, merely that it identifies mass-inhomogeneities $\delta m$ therein 
concentrated over a limited azimuthal extend, i.e., $\delta \varphi/2\pi <<1$ in each ring of sufficiently small width $\delta r$ (Fig. 1). 
To be precise, consider a spiral wave pattern in the orbital plane $(x=r(\varphi)\cos\varphi, y=r(\varphi)\sin\varphi)$ in polar coordinates $(r,\varphi)$ satisfying $\left|dr(\varphi)/d\varphi\right|\ge \kappa>0$ for all $\varphi$. (Here, $\kappa=0$ is
excluded, as $dr(\varphi)/d\varphi=0$ corresponds to a circle.) Then $\delta \varphi \le \delta r/\kappa$, and hence
$\delta\varphi/2\pi<<1$ for $\delta r$ sufficiently small.

We next consider a Lagrangian description, in which the accretion disk is partitioned in rings. A ring 
located at a radius $r$ has a mass 
\begin{eqnarray}
\Delta m(r)\equiv \sigma(r) M
\end{eqnarray}
 and a radial width  $l(r)$.   It is naively anticipated that  $l(r)$ is on the order of the vertical scale height $H(r)$ of the disk.   
For a thin ring of radius $r$ rotating at a frequency $\Omega$ in the $(x,y)$ plane, the non-vanishing components of the quadrupole moment tensor, 
\begin{eqnarray}
I^{ij}=\int_Vd^3x\rho(t,\vec{x})x^ix^j,
\end{eqnarray}
where the integration is over the ring's volume,  are given, to a good approximation, by 
\begin{eqnarray}
\begin{array}{rll}
I^{xx}&=\frac{1}{2}\Delta mr^2(1+\xi\cos2\Omega t),\\
I^{yy}&=\frac{1}{2}\Delta mr^2(1-\xi\cos2\Omega t),\\
I^{xy}&=-\frac{1}{2}\xi\Delta mr^2\sin2\Omega t,
\end{array}
\end{eqnarray}
in terms of the dimensionless parameter $\xi$ that quantifies the quadrupole mass inhomogeneity, whereby  $\xi=0$ for
an axi-symmetric ring, $\xi=0.5$ for a ring having a density $\rho(r,\varphi)=\rho_0\cos^2\varphi$, and $\xi=1$ for two identical point masses in circular orbit around
their center of mass.

For simplicity, the dynamics of the ring and its gravitational wave emission are computed in the Newtonian limit.   This is a reasonable approximation far enough from the ISCO.   However, as shown below, the gravitational wave power is maximized near the ISCO, where absorption and scattering of gravitational waves and other GR effects are at play.  Thus, our analysis should be treated as a first approximation. 

Orbital angular momentum is lost both in viscous torques and gravitational radiation, leading to an inward radial drift of the ring. For  $\xi>0$ the gravitational wave power of the ring, $\Delta L_{GW}(r)$, is assumed to be dominated by $l=m=2$ mode described by the quadrupole formula,
\begin{equation}
\frac{dE_{GW}}{dt}=\frac{1}{5}\langle\frac{d^3{\cal I}_{jk}}{dt^3}\frac{d^3{\cal I}^{jk}}{dt^3}\rangle,\label{dEgw/dt}
\end{equation}
where ${\cal I}^{jk}=I^{jk}-\frac{1}{3}\delta^{jk}\delta_{lm}I^{lm}$, i.e., 
\begin{equation}
\Delta L_{GW}(r)\simeq \frac{32}{5}\xi^2\left(\frac{M}{r}\right)^5\left(\frac{\Delta m}{M}\right)^2=\frac{32}{5}\xi^2\left(\frac{M}{r}\right)^5\sigma^2(r).
\label{gw-ring}
\end{equation}
With our assumption on $m=2$ dominance, we focus on rings that are luminous in gravitational waves and mechanisms of non-axisymmetric instabilities for which low order modes tend to be unstable first. Furthermore, the gravitational wave output of a 
collection of rings that are each $m=2$ unstable will be the sum of (\ref{gw-ring}), each at its own gravitational wave frequency
defined by twice the Keplerian frequency (\ref{EQN_OMK}).

\begin{figure}[h]
\centerline{\includegraphics[width=13cm]{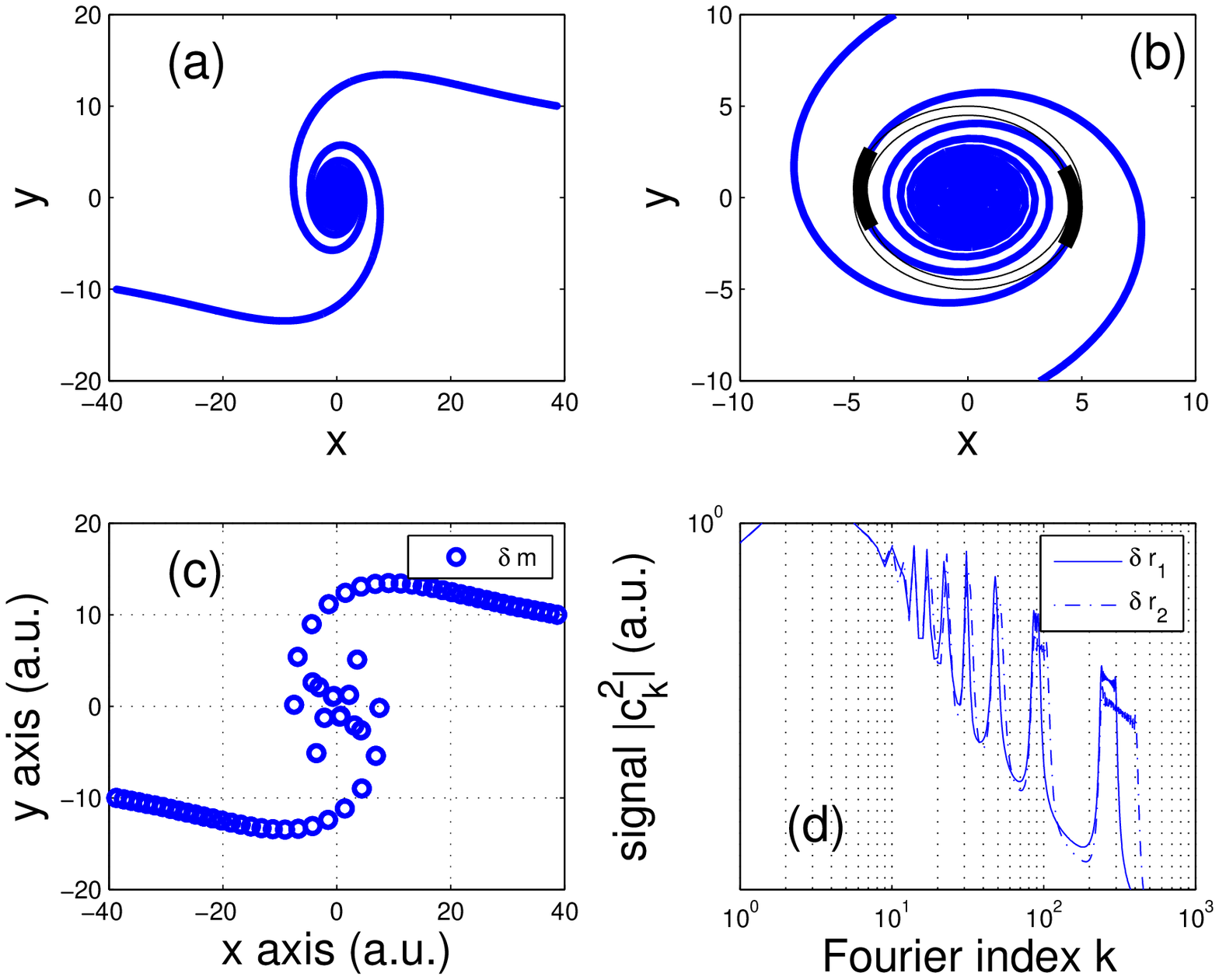}}
\caption{(a) A spiral density wave pattern in a disk. (b) Identification of over-dense regions (thick black) in an
annular region $4.5 < r < 5$ (thin black circles) that are of finite angular extend $\delta \varphi/2\pi<<1$. 
(a) Leading order approximation of the over-dense regions in (b) by local mass-inhomogeneities $\delta m$, here $2\times 32$ in number following a grid with 32 annular rings. (d) By their individual and distinct Keplerian angular frequencies, the quadrupole emission spectra of each $\delta m$ in (c) are non-overlapping. Expressed in terms of the modulus squared of the associated Fourier coefficients $c_k$, accretion broadens the emission of each, shown in (d) for two rates of accretion corresponding to different radial 
migrations $\delta r_2<\delta r_1<0$.}
\label{fig:0}
\end{figure}

An illustrative example is a spiral wave pattern in Fig. \ref{fig:0}. Overlaying our discretization of the plane of the disk in
annular rings, this pattern introduces over-dense regions in each ring. 
Each ring hereby assumes a $m=2$ quadrupole moment inherited from the underlying spiral wave,
here represented by a pair of mass-inhomogeneities $\delta m$ in each ring. Due to rotation, each $\delta m$ 
emits quadrupole gravitational wave emission at twice the local Keplerian frequency. 
Subject to accretion, the tidal wave gradually tightens. As the $\delta m$ gradually migrate inwards, these quadruple emissions
broaden in frequency, each taking their own window in the gravitational wave spectrum. By Parseval's theorem, the total luminosity
from the spiral wave is the sum of the luminosities in each ring.

It may be appreciated that our focus on quadrupole gravitational wave emissions is a leading order approach, that ignores the possibility of emissions from higher order mass moments. A formalism for calculating the full spectrum of gravitational wave emission from multiple mass moments in an annular ring is given in \cite{bro06}.

For gravitational wave driven accretion, the loss of orbital angular momentum in gravitational waves causes a radial displacement $dr$ of the ring on a time
\begin{equation}
dt_{GW}\simeq \frac{d(\Delta E)}{\Delta L_{GW}}=\frac{5}{64} \left(\frac{r}{M}\right)^3\xi^{-2}\sigma(r)^{-1}dr,\label{t_gw}
\end{equation}
where $\Delta E=-\Delta m/2r$ is the energy of the ring  in the Newtonian limit. 
Viscous torques acting on the ring cause a radial drift $dr$ on a time \citep{pri81}
\begin{equation}
dt_\nu=\frac{2rdr}{3\nu}=\frac{\sqrt{8}}{3\alpha}\left(\frac{r}{H}\right)^2\left(\frac{r}{M}\right)^{1/2}dr,
\end{equation}
where the alpha prescription is adopted to express the kinematic viscosity in terms of the vertical scale height $H(r)$ and the $\alpha$ parameter as $\nu=\alpha c_s H=\alpha \Omega H^2/\sqrt{2} $,
using $c_s=\Omega H/\sqrt{2} $ between the sound speed and angular velocity $\Omega$.
The two times are equal at $r=r_b$ given implicitly by   
\begin{equation}
\frac{r_b^{1/2}H_b^2}{M^{5/2}\sigma_b}=\frac{128\sqrt{2}}{15}\frac{\xi_b^2}{\alpha},
\label{rb}
\end{equation}
here the subscript $b$ denotes values at $r_b$.

For viscous driven accretion, i.e., at radii where angular momentum losses are dominated by viscous torques, the surface density of the disk is given approximately by $\Sigma(r)=\dot{m}/(3\pi\nu)$  for accretion rate $\dot{m}$ \cite{pri81}.   The width of the ring can be identified with the scale height of the disk, hence we set $l(r)=H(r)$. Then, the mass of the ring can be approximated as
\begin{equation}
\sigma(r)=2\pi r l(r) \Sigma(r)=\frac{\sqrt{8}}{3\alpha}  \left(\frac{r}{H}\right)\left(\frac{r}{M}\right)^{3/2}\dot{m}.\label{sig}
\end{equation}
Combining Equations (\ref{rb}) and (\ref{sig}) yields
\begin{equation}
\left(\frac{H_b}{r_b}\right)^{2}\left(\frac{H_b}{M}\right)=\frac{512}{45}\frac{\xi_b^2}{\alpha^2}\dot{m}.
\end{equation}
At $r>r_b$ viscous stresses contribute the dominant torque, whereas at $r_{isco}< r< r_b$ the torque is dominated by gravitational radiation. Semi-analytic models of neutrino cooled disks \cite{pop99,che07} indicate that $H/r$ changes from about $0.1$ near the ISCO to $0.2$ at $r=10M$ to $0.45$ at $r=100 M$, with relatively weak dependence on $a$ and $\dot{M}$ for the conditions relevant to our analysis.  We shall therefore adopt for illustration $H(r)=\eta r$,  $\eta=0.1\eta_{-1}$, in the inner disk region.   Denoting $\alpha_{-1}=\alpha/0.1$,  we obtain a rough estimate for the transition radius:
\begin{equation}
\frac{r_b}{M}=\frac{512\xi_b^2}{45\alpha^2 \eta^3}\dot{m}\simeq 6\xi_b^2\eta_{-1}^{-3} \alpha_{-1}^{-2}\left( \frac{\dot{M}}{1\ M_\odot\ s^{-1}}\right).
\label{rb/M}
\end{equation}
We thus arrive at (\ref{rb}) and (\ref{rb/M}) constraining $r_b$. 

In the {\em outer region} $r>r_b$ the mass profile is given by  Equation (\ref{sig}).  Combining with Equation (\ref{gw-ring}) yields 
\begin{equation}
\Delta L_{GW}(r>r_b)=\frac{256}{45}\frac{\dot{m}^2}{\alpha^2 \eta^2}\left(\frac{M}{r}\right)^2\xi^2(r)  \label{gw-ring-out}.
\end{equation}
In the {\em inner region} $r_{isco}<r<r_b$ the mass profile of the rings is determined from the relation $\sigma dr/l=\dot{m}d t_{GW}$.
Then, by employing Equation (\ref{t_gw}) one has 
\begin{equation}
\sigma(r<r_b)=\frac{\sqrt{5}}{8}\eta^{1/2}\dot{m}^{1/2}\xi^{-1}\left(\frac{r}{M}\right)^{2},\label{sig-gw}
\end{equation}
and $\Delta L_{GW}(r<r_b)=\dot{m}\eta M/2r$, upon substituting Equation (\ref{sig-gw}) into (\ref{gw-ring}).

The total luminosity, $L_{GW}$, is the sum over the contributions of all the rings, each radiating at their own frequency as indicated above.
Approximating the sum by an integral we can write: $L_{GW}\simeq \int_{r_{isco}}^{r_{out}} \Delta L_{GW}(r)dr/l$.  In the parameter regime of inefficient emission,
\begin{eqnarray}
r_b<r_{isco},
\label{EQN_A}
\end{eqnarray}
the mass profile given by Equation (\ref{sig}) holds everywhere and hence, using Equation (\ref{gw-ring-out}), we further have
\begin{equation}
\frac{L_{GW}}{\dot{M}c^2}\simeq \int_{r_{isco}}^{r_{out}}\frac{256}{45}\frac{\xi^2 M^2 dr}{\alpha^2 H^3}\dot{m}\simeq\frac{128}{45}\frac{<\xi^2>\dot{m}}{\alpha^2 \eta^3z^2}
\end{equation}
i.e.,
\begin{equation}
\frac{L_{GW}}{\dot{M}c^2}\simeq 1.5\frac{<\xi^2>}{\eta_{-1}^3\alpha_{-1}^2z^2}\left( \frac{\dot{M}}{1\ M_\odot\ s^{-1}}\right)\label{Lgw-ineffic}
\end{equation}
for $r_{out}>>r_{isco}$.  Here $<\xi^2>=2r_{isco}^2\int r^{-3}\xi^2(r)dr$, and $z=r_{isco}/M$.  The parameter regime
\begin{eqnarray}
r_b>r_{isco}
\label{EQN_B}
\end{eqnarray}
is different.  In this regime Equations (\ref{sig-gw})  and (\ref{gw-ring}) yield
\begin{equation}
\frac{L_{GW}}{\dot{M}c^2}\simeq   \frac{1}{2z}\label{Lgw-effic}
\end{equation}
for $r_b>>r_{isco}$.  Efficient emission occurs at accretion rates for which Equation (\ref{EQN_B}) is satisfied. With aforementioned definition of $z$,
Equation (\ref{rb/M}) gives the corresponding criterion
\begin{eqnarray}
\dot{M}\ge0.15 \eta_{-1}^3<\xi^{2}>^{-1}\alpha_{-1}^2z M_\odot s^{-1}
\label{EQN_M12}
\end{eqnarray}
whenever a mass inhomogeneity $\xi$, assumed to originate at larger radii, survives all the way to the ISCO. 
If it dissipates at some radius $r_{diss}>r_{isco}$, then $r_{isco}$ should be replaced by $r_{diss}$, 
thus raising the minimum accretion rate on the right hand side of (\ref{EQN_M12}). 
 \begin{figure}[h]
\centerline{\includegraphics[width=12.5cm]{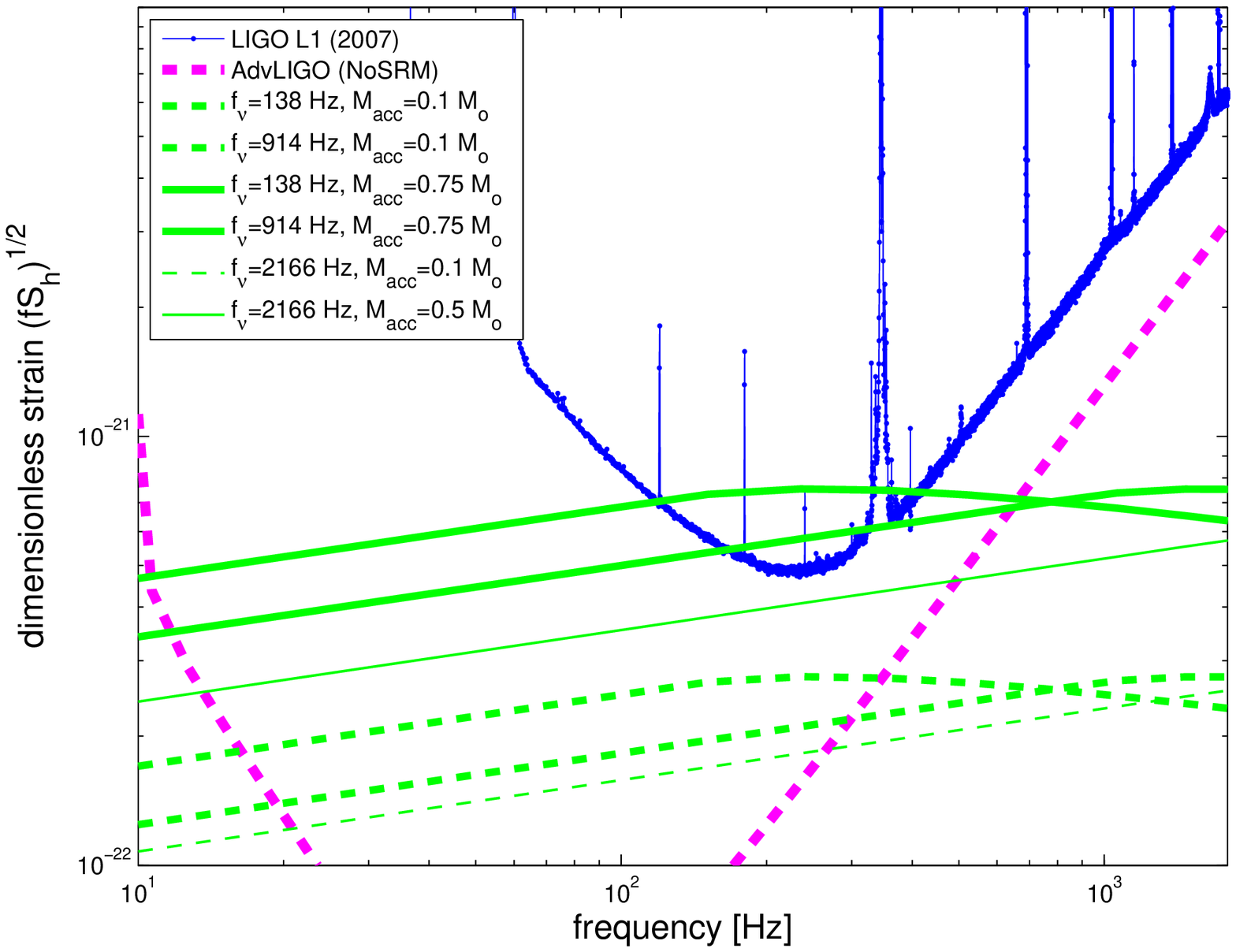}}
\caption{Characteristic strain amplitude $h_{char}(f)$ of broadband quadrupole emission in 
accretion flows around a stellar mass black holes of $M=10M_\odot$ at
a fiducial distance of $D=100$ Mpc. The vertical distance to the 
dimensionless strain noise $h_n=\sqrt{fS_h}$ of LIGO S5 (2007) represents the theoretical limit of signal-to-noise ratio in a 
matched filtering detection. The different lines correspond to different values of $f_b$ and $M_{a}=\dot{m}\tau$, as indicated. Thick 
lines correspond to the maximally efficient parameter regime $f_b<f_{isco}$ in Eq. (\ref{h_char-tot}); thin lines correspond to $f_b>f_{isco}$.}
\label{fig:1}
\end{figure}
\begin{figure}[h]
\centerline{\includegraphics[width=12.5cm]{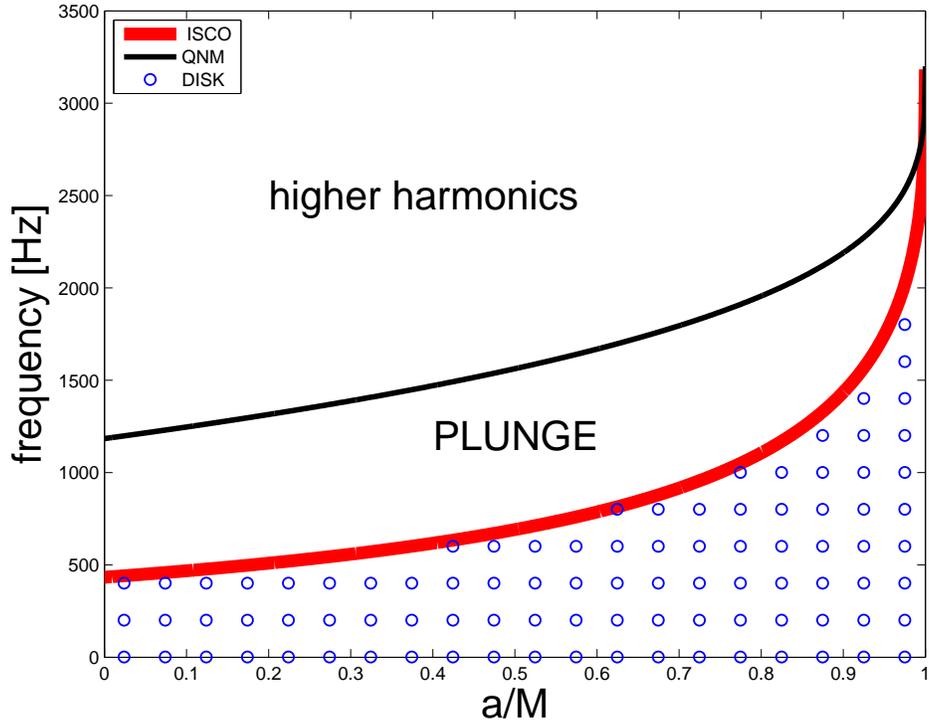}}
\caption{Schematic overview of quadrupole gravitational radiation from  hyper-accreting black holes. Wave patterns in accretion flow may produce broadband emission (circles), whereas fragmentation \citep{pir07} or non-axisymmetric waves in a torus about the ISCO \citep{van03} (red curve) may produce chirps. Relatively high frequency emission may derive from Quasi-Normal Mode (QNM) ringing in the black hole event horizon (black curve).}
\label{fig:2}
\end{figure}

\begin{figure}[h]
\centerline{\includegraphics[width=12.5cm]{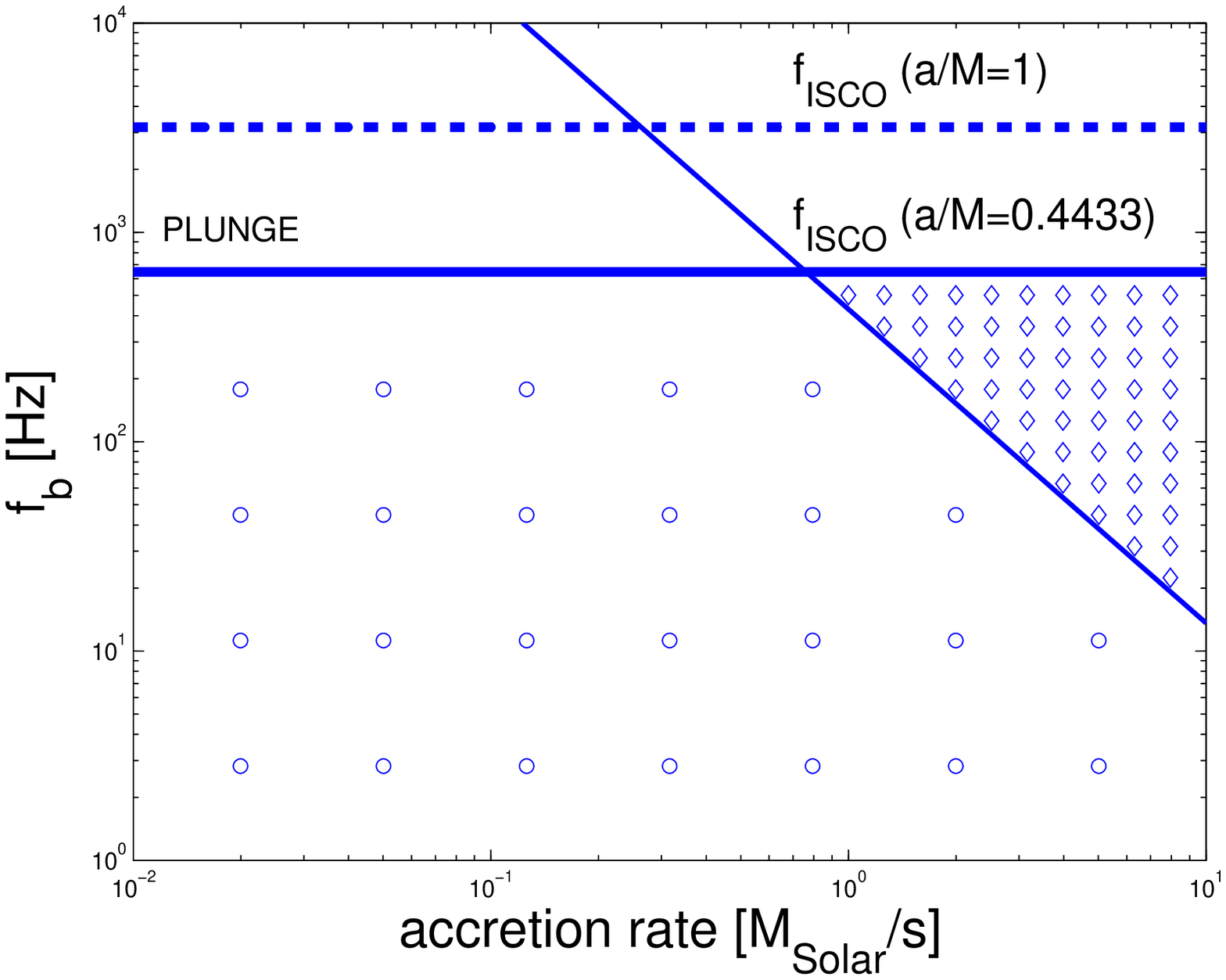}}
\caption{A plot of $f_b$ versus accretion rate $\dot{M}$, for the canonical choice of parameters adopted
in Eq. (\ref{f_b}).  The horizonthal lines delineate the ISCO frequence for two choices of the spin parameter,
as indicated. The region enclosed above the $f_b$ line and below the $f_{isco}$ line (diamonds) indicates
the regime of maximum efficiency of gravitational wave emission.  The circles indicate the permitted frequency 
range of less efficient emission, though still relevant for detection by advanced gravitational wave detectors, as indicated 
by the thin lines in Fig. \ref{fig:1}.}
\label{fig:3}
\end{figure}

For a matched filtering detection method, the relevant quantity is the amplitude that takes into account the square root of the associated number of wave periods. In the frequency domain, the corresponding quantity is the characteristic strain amplitudel  \citep[e.g.,][]{cut02}:  
\begin{eqnarray}
h_{char}(f)=\frac{\sqrt{2}}{\pi D}\sqrt{\left|\frac{\Delta E}{\Delta f}\right|}.
\end{eqnarray}
Consider now coherent emission episode of duration $\tau$.    In the accretion model outlined above, $\tau$ corresponds to the period during which a steady disk pattern, excited by some instability, survives, satisfying
\begin{eqnarray}
t_{GW}<\tau<T,
\end{eqnarray}
where $T$ is the total accretion time. The energy radiated over time $\tau$ from an annular region located between $r$ and $r_{out}>>r$, assuming $r_b>r>r_{isco}$, is
\begin{eqnarray}
\Delta E_{rad}=\tau\int_r^{r_{out}}\Delta L_{GW}(r')dr'\simeq \dot{m}\tau\left(\frac{M}{2r}\right)\left(1-\frac{r}{2r_b}\right).
\label{EQN_DE}
\end{eqnarray}  
We remark Eqn. (\ref{EQN_DE}) is independent of $\xi$, that governs the gravitational wave luminosity
Eqn. (\ref{gw-ring}). As such, $\xi$ governs the time scale over which $\Delta E_{rad}$ is realized.
A small (large) $\xi$ has a low (high) luminosity in GWs, whereby $\Delta E_{rad}$ takes a long (short) time to 
come to fruition. The net result, however, is always the same $\Delta E_{rad}$. 

The energy $\Delta E_{rad}$ is emitted over the frequency interval $\Delta f=f(r)-f(r_{out})=\Omega(r)/\pi-\Omega(r_{out})/\pi\simeq \pi^{-1}M^{-1}(M/r)^{3/2}$, for the assumed Keplerian rotation.    Thus, the characteristic strain amplitude is
$h_{char}(f>f_b)=\frac{\sqrt{2}}{\pi D}\sqrt{\left|\frac{\Delta E_{rad}}{\Delta f}\right|}$, i.e.,
\begin{equation}
h_{char}(f>f_b)=\frac{M}{\sqrt{2\pi}D}\sqrt{\frac{\dot{m}\tau}{M}}(\pi M f)^{-1/6}\sqrt{2-\left(\frac{f_b}{f}\right)^{2/3}},
\end{equation}
denoting $f_b=f(r_b)$.  In the outer region $r>r_b$ we employ Eq. (\ref{gw-ring-out}) to obtain $\Delta E_{rad}\simeq\tau \Delta L_{GW}\propto (r/M)^{-2}$, emitted over the 
frequency interval $\Delta f\simeq f(r)$.  Thus, $h_{char}(f<f_b)\propto f^{1/6}$.  Careful calculations yield:
\begin{equation}
h_{char}(f)=\kappa \left\{\begin{array}{ll}
          \left(\frac{f}{f_b}\right)^{1/6}
 & \mbox{($f <f_b$)}\\
        \left(\frac{f}{f_b}\right)^{-1/6}\sqrt{2-\left(\frac{f_b}{f}\right)^{2/3}} & \mbox{($f_{isco}>f > f_b$)}\end{array} \right. 
\label{h_char-tot}
\end{equation}
with $\kappa$ scaled to distances $D_{2}= \left({D}/{100\,\mbox{Mc}}\right)$ and masses $M_1=M \,10M_\odot$ satisfying
\begin{equation}
\kappa=\sqrt{8}\times10^{-22}D_{2}^{-1} M_1^{1/3}\left(\frac{\dot{m}\tau}{0.1M_\odot}\right)^{1/2}
\left(\frac{f_b}{1000  Hz}\right)^{-1/6}.
\label{kappa}
\end{equation}
Here $f_{isco}=f(r_{isco})$ is the frequency in quadrupole GW emission of matter at the isco.  Eq. (\ref{h_char-tot}) applies also in the regime $f_b>f_{isco}$, whereby $h_{char}=\kappa (f/f_b)^{1/6}$, $f\le f_{isco}$.

Fig. \ref{fig:1} displays a plot of Eq. (\ref{h_char-tot}) for different values of $f_b$ and $\dot{m}\tau$.
It shows the characteristic  strain amplitude peaks at the frequency $f_b$, given by
\begin{equation}
f_b=430\eta_{-1}^{9/2}\left(\frac{\alpha_{-1}}{\xi}\right)^3\left(\frac{M}{10M_\odot}\right)^{-1}\left( \frac{\dot{M}}{1\ M_\odot\ s^{-1}}\right)^{-3/2}\quad {\rm Hz},
\label{f_b}
\end{equation}
where Eq. (\ref{rb/M}) has been used. The condition $r_b>r_{isco}$ implies $f_b<f_{isco}$, where for a $10 M_\odot$ black hole $430\ {\rm Hz}<f_{isco}<3000\ {\rm Hz}$, depending on the spin parameter $a/M$ of the black hole,  with $f_{isco}=1600$ Hz for $a/M=0.95$ as an example.  

Fig. \ref{fig:2} shows the frequency range relevant to gravitational wave emission from hyper-accreting black holes;
Fig. \ref{fig:3} exhibits a plot of $f_b$ versus $\dot{M}$, elucidating the dependence on the spin parameter.

\section{Conclusion and Outlook}

Fig. \ref{fig:1} indicates a window of extended gravitational wave emission, that may arise from long-lived quadrupole mass inhomogeneities in 
hyper-accretion flows. As a result of accretion, these inhomogeneities will produce chirps, as opposed to constant frequency emission such as may
be produced by a pulsar. In deriving our scaling relation (\ref{f_b}) for the peak luminosity in gravitational waves, as illustrated in Fig. 2, we display 
a radiatively efficient and inefficient parameter regime (\ref{EQN_A}) and, respectively, (\ref{EQN_B}), depending on the viscosity and accretion rate in 
the accretion flow. 

Broadband emission in chirps can be searched for broadband spectra extracted from noisy time series by a Time Sliced Matched Filtering (TSMF, \cite{van11}). In \cite{van14}), some 8.6 million templates were used to densely cover the intended range in frequency and frequency time rate-of-change. In the application to gravitational wave searches, exhaustive searches of this kind require massively parallel computing and a relatively accurate determination of the True Time Of Onset (TTOO) of the supernova, which may be estimated by extrapolation of its optical light curve, that may be obtained from dedicate supernova surveys of the Local Universe \citep{heo15}.

A LIGO-Virgo or KAGRA detection of the proposed broadband emissions promises entirely novel constraints on the viscosity and accretion rates in core-collapse supernovae, not currently accessible by existing electromagnetic observations. Considerable gain in sensitivity can be attained by coherent network analysis of the four
detector sites of these observatories (e.g. \cite{wen05,hay15}). Detection promises unique insight into the different phases of birth 
and evolution of stellar mass black holes. Probes of broadband extended gravitational emission such as illustrated by spiral waves in Fig. \ref{fig:0} may, 
for instance, identify a critical accretion rate for hyper-accretion \citep{glo14}, the end of which leaves a near-extremal black hole as the initial condition to
pronounced ISCO waves during prompt GRB emission in subsequent phase of black hole spin down \citep{van15}.

{\bf Acknowledgments.} The authors thank the referee for several constructive comments to the manuscript. 
A. Levinson acknowledges support  by a grant from the Israel Science Foundation no. 1277/13. M.H.P.M. van Putten acknowledges
partial support from a Sejong University Faculty Research fund.

\label{lastpage}

\end{document}